\documentclass{article}

\usepackage{arxiv}

\usepackage[utf8]{inputenc} 
\usepackage[T1]{fontenc}    
\usepackage{hyperref}       
\usepackage{url}            
\usepackage{booktabs}       
\usepackage{nicefrac}       
\usepackage{microtype}      
\usepackage{amsfonts}       
\usepackage{lipsum}         
\usepackage{graphicx}
\usepackage{natbib}
\usepackage{doi}
\usepackage{graphicx}
\usepackage{xcolor}
\usepackage{hyperref}
\usepackage{multirow}
\usepackage{amsmath}
\usepackage{bm}

\usepackage{algpseudocode}
\usepackage[english,onelanguage,ruled]{algorithm2e}

\def\bSig\mathbf{\Sigma}

\newcommand{\Tr}{\mbox{tr}}

\newcommand{\sumas}{\sum^n_{i=1}}

\newcommand{\ii}{i=1,\ldots,n}
\newcommand{\sumasp}{\sum^p_{j=1}}

\newcommand{\Y}{\mathbf{Y}}

\newcommand{\yp}{\mathbf{y}}
\newcommand{\by}{\mathbf{y}}
\newcommand{\x}{\mathbf{x}}

\newcommand{\Z}{\mathbf{Z}}

\newcommand{\ba}{\mathbf{b}}

\newcommand{\bD}{\mathbf{D}}

\newcommand{\X}{\mathbf{X}}
\newcommand{\y}{\mathbf{y}}

\newcommand{\A}{\textrm{A}}
\newcommand{\balpha}{\mbox{${\bm \alpha}$}}
\newcommand{\bmu}{\mbox{${\bm \mu}$}}

\newcommand{\bSigma}{\mbox{${\bm\Sigma}$}}
\newcommand{\bepsilon}{\mbox{${\bm \epsilon}$}}
\newcommand{\bLambda}{\mbox{${\bm \Lambda}$}}
\newcommand{\bbeta}{\mbox{${\bm \beta}$}}
\newcommand{\btheta}{\mbox{${\bm \theta}$}}

\newcommand{\brho}{\mbox{${ \bm \rho}$}}

\title{The use of the EM algorithm for regularization problems in  high-dimensional linear mixed-effects models}

\author{ \href{https://orcid.org/0000-0002-9573-8424}{\includegraphics[scale=0.06]{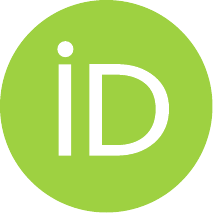}\hspace{1mm}Daniela C. R.~Oliveira}\\ 
	Department of  Mathematics and Statistics\\
	Federal University of Sao Joao del-Rei\\
	MG, Brazil \\
	\texttt{daniela@ufsj.edu.br} \\
	\And
	\href{https://orcid.org/0000-0002-5724-8918}{\includegraphics[scale=0.06]{orcid.pdf}\hspace{1mm}Fernanda L.~Schumacher} \\
	Division of Biostatistics\\
 College of Public Health\\
	The Ohio State University\\
	OH, U.S.A. \\
	\texttt{schumacher.313@osu.edu} \\
 \AND
	\href{https://orcid.org/0000-0002-7239-2459}{\includegraphics[scale=0.06]{orcid.pdf}\hspace{1mm}V\'{\i}ctor H.~Lachos} \\
	Department of Statistics\\
	University of Connecticut\\
	CT, U.S.A. \\
	\texttt{hlachos@uconn.edu} \\
}

\renewcommand{\shorttitle}{EM algorithm for regularization problems in high-dimensional LMMs}

\hypersetup{
pdftitle={EM algorithm for regularization problems in high-dimensional LMMs},
pdfsubject={stat.ME},
pdfauthor={Daniela C. R.~Oliveira, Fernanda L.~Schumacher, Victor H.~Lachos},
pdfkeywords={EM algorithm, High-dimensional data,  Mixed-effects models, R package glmnet,  Regularized variable selection methods},
}

\date{July 31, 2023}

\begin{document}
\maketitle

\begin{abstract}
	The EM algorithm is a popular tool for maximum likelihood estimation but has not been used much for high-dimensional regularization problems in linear mixed-effects models. In this paper, we introduce the \texttt{EMLMLasso} algorithm, which combines the EM algorithm and the popular and efficient \textsf{R} package \textbf{glmnet} for Lasso variable selection of fixed effects in linear mixed-effects models. We compare the performance of our proposed \texttt{EMLMLasso} algorithm with the one implemented in the well-known \textsf{R} package \textbf{glmmLasso} through the analyses of both simulated and real-world applications. The simulations and applications demonstrated good properties, such as consistency, and the effectiveness of the proposed variable selection procedure, for both $p < n$ and $p > n$. Moreover, in all evaluated scenarios, the \texttt{EMLMLasso} algorithm outperformed \texttt{glmmLasso}. The proposed method is quite general and can be easily extended for ridge and elastic net penalties in linear mixed-effects models.
\end{abstract}



\section{Introduction}\label{Intro}

The linear mixed-effects models (LMM) are a class of statistical models used to describe the relationship between the response and covariates based on clustered or longitudinal data\citep{Laird:1982}. Such data are becoming increasingly popular in many subject-matter areas, especially genetics, health, finance, ecology, and image processing. Selecting the best LMM for these data is crucial.  An important issue arises when the number of predictors $(p)$ is high compared to the number of observations $(n)$, i.e., $p>n$. This is broadly
known as high-dimensional variable selection \citep{Buhlmann2014}. Even with the advancement of computational, statistical, and technological tools, the selection of fixed effects in LMM under high dimensionality is still a challenge. In this paper, we focus on the selection of these effects with a special focus on the Least Absolute Shrinkage and Selection Operator (Lasso) \citep{tibshirani1996regression}.

There are many statistical methods proposed for variable selection. Among them, a popular class of methods is variable selection via regularization, also known as penalized variable selection. This class of methods has the key advantage of simultaneously selecting important variables. In the context of fixed effects selection in LMM, \cite{Schelldorfer2011} and \cite{Groll2022} estimated the parameters based on L1-penalization  that maximize the penalized log-likelihood (PML) function using computational methods. By considering the random effects as missing values in the LMM framework, \cite{ROHART2014} proposed an L1-penalization on the fixed effects coefficients of the resulting log-likelihood, with the optimization problem solved via a multicycle expectation-maximization (EM) algorithm. 

\cite{Ghosh2018} considered the selection of important fixed-effect variables in LMM along with PML estimation of both fixed and random-effect parameters based on general non-concave penalties. \cite{Ghosh2021} proposed a generalized method-of-moments approach to select fixed effects in the presence of a correlation between the model error and the covariates. Recently, \cite{AlabisoShang2022} proposed a conditional thresholded partial correlation algorithm to select fixed effects in LMM. There are several proposals for selecting fixed and random effects simultaneously in LMM. However, we highlight the works that use the selection of fixed effects, and we chose to use the publicly available \textsf{R} package \textbf{glmmLasso} \citep{Groll2022} for comparison purposes. The \texttt{glmmLasso}  is a gradient ascent algorithm designed for generalized linear mixed models, which incorporates variable selection by L1-PML estimation. In a final re-estimation step, a model that includes only the variables corresponding to the non-zero fixed effects is fitted by simple Fisher scoring.

The goal of this paper is to propose a new approach for fixed effects selection that combines the popular EM algorithm with PML estimation using the Lasso penalty. The PML step is performed via \texttt{glmnet} \citep{friedman2010regularization},  which is an efficient and reliable \textsf{R} package publicly available that fits a generalized linear model via PML, and the regularization path is computed for the Lasso, ridge or elastic net penalty at a grid of values for the tuning (regularization) parameter. In addition, the  Bayesian Information Criterion (BIC) is used to determine the optimal tuning parameter. In this work, we focus on the Lasso penalty, which is called the \texttt{EMLMLasso} algorithm, but extensions to include other penalties are straightforward. The final model, which includes only the variables corresponding to the non-zero fixed effects, can be fitted using standard \textsf{R} packages such as \textbf{lme4} \citep{bates_lme4} or \textbf{skewlmm} \citep{schumacher2020scale}.

\begin{figure}[htb]
\centering
\includegraphics[width=12cm,height=12cm]{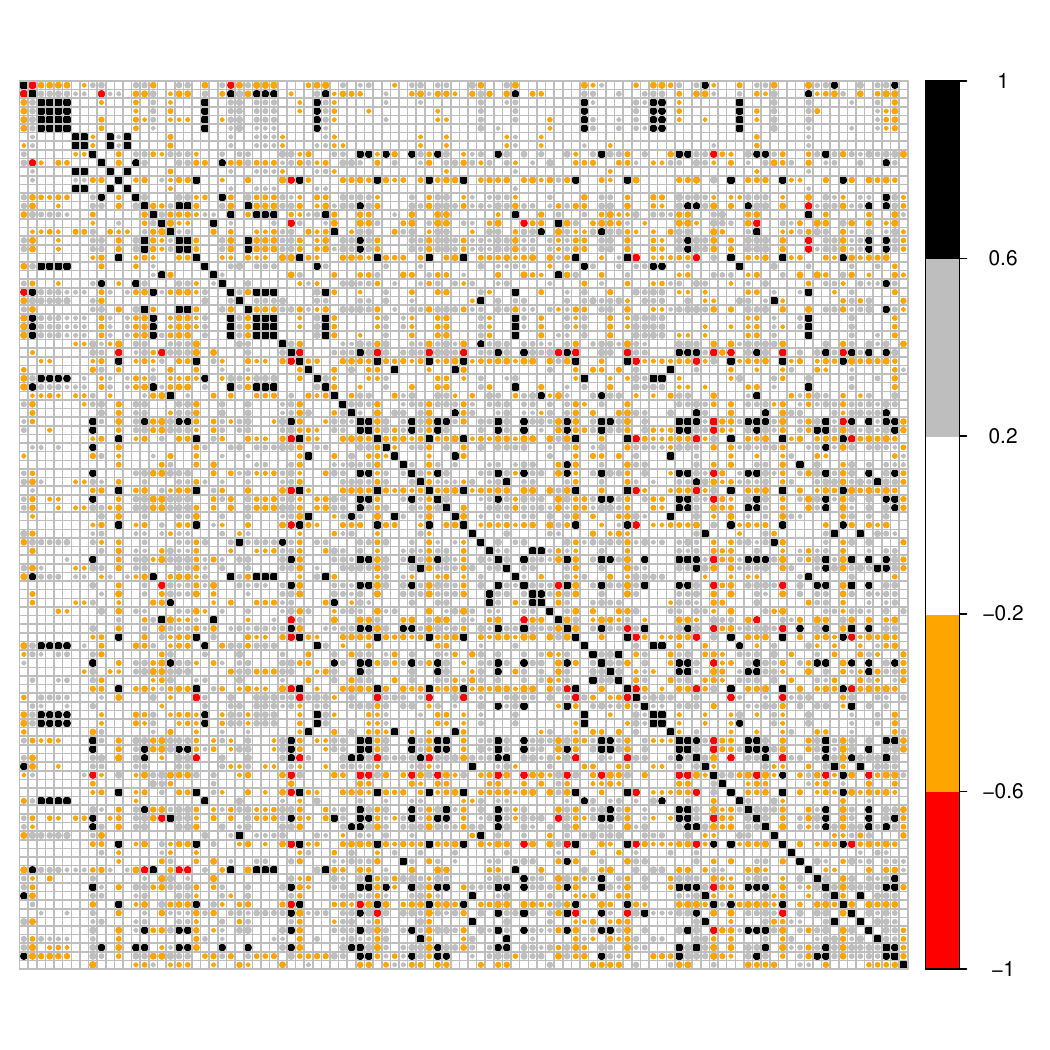}
\caption[FIGURE 1]{Correlation plot of covariates for riboflavin dataset, with $n=71$ and $p=101$ ($p>n$). }
\label{Ap2Introd}
\end{figure}

Our motivating datasets in this study are two folded: (I) The first is the well-known Framingham cholesterol study, where selected variables $(p<n)$ can explain the level of cholesterol, which is a risk factor for the evolution of cardiovascular diseases, we also evaluate the effectiveness of the algorithms by including three simulated variables; (II) In the second real data set, we looked for relevant genes $(p>n)$ that can increase the production of the riboflavin (vitamin B2) of \textit{bacillus subtilis}, a bacterium found in the human digestive tract. The correlation between the covariates can be seen in Figure \ref{Ap2Introd} in the colored spots off the main diagonal 
The riboflavin dataset contains $p=101$ covariates representing the log of the expression level of $100$ genes and the time with a total of $n=71$ observations.

The remainder of the paper is organized as follows. In Section \ref{backgraund}, we introduce the general linear regression model, the Lasso penalty, regularization problems, and the EM algorithm. In Section \ref{sec:model}, we introduce the linear mixed-effects models and present \texttt{EMLMLasso} algorithm. In Section \ref{sec:experiments}, we show the results of the simulation experiments, demonstrating the effectiveness of the proposed algorithm. In Section \ref{sec:realdata}, we apply the proposed algorithm to two real datasets. Finally, we present some concluding remarks and perspectives for future research in Section \ref{sec:conclusions}.

\section{Preliminaries } \label{backgraund}

 We begin our exposition by defining the notation and presenting some basic concepts which are used throughout the development of our methodology. As is usual in probability theory and its applications, we denote a random variable by an upper-case letter and its realization by the corresponding
lower case and use boldface letters for vectors and matrices.  $\mathrm{N}_p(\bmu, \bSigma)$ denotes the $p$-variate normal distribution with mean vector $\bmu$ and variance-covariance matrix $\bSigma$. When $p=1$, we drop the index $p$. $\A^{\top}$ denotes the transpose of $\A$ and $\Tr(A)$ is a trace of $A$, i.e., sum of elements on the main diagonal of $A$.   $\mathbf{I}_ p$ denotes the $p
\times p$ identity matrix.

\subsection{The linear model}
First, consider the general linear regression model
\begin{eqnarray}\label{ML1}
\Y_i=\X_i\bbeta+\bepsilon_i,\,\,\ii,
\end{eqnarray}
where  $\Y_i$ is a  $n_i$-vector of continuous outcome, $\X_i$ is a $n_i\times p$ matrix of fixed predictors, $\bbeta=(\beta_1,\ldots,\beta_p)^{\top}$ is a $p$-vector of unknown regression parameters, and $\bepsilon_i$ are independent and identically normally distributed random vectors with mean $\bf{0}$ and variance-covariance matrix $\sigma^2\mathbf{I}_{n_i}$;  $\mathbf{I}_ p$ denoting the $p
\times p$ identity matrix.
Let $\X$ be the $N\times p$ matrix $\X=(\X_1,\ldots \X_n)^{\top}$ and $\Y$ the $N\times 1$ vector  $\Y=(\Y^{\top}_1,\ldots,\Y^{\top}_n)^{\top}$, with $N=\sumas n_i$. 
The intercept is omitted in the model for simplicity, and all predictors and the response variable are assumed to be standardized (i.e., zero mean and unit variance).

\subsection{Lasso and regularization problems}

Lasso is one of the most popular methods in high-dimensional data analysis. It allows for simultaneous estimation and variable selection and has efficient
algorithms available \citep{buhlmann2011statistics}. The Lasso estimator of parameters in the model (\ref{ML1}) is:
\begin{eqnarray}\label{PLSequation}
\widehat{\bbeta}(\lambda)&=&\underset{\bbeta}{\arg\min}\left\{(\Y-\X\bbeta)^{\top}(\Y-\X\bbeta)+\lambda\, \Psi_p(\bbeta)\right\},
\end{eqnarray}
where  $\Psi_p(\bbeta)=||\bbeta||_1=\sumasp|\beta_j|$ ($L_1$ norm) is the Lasso penalty on the parameter size, and $\lambda>0$ controls the amount of regularization. 

In practice, $\lambda$ depends on the data, and its optimal value can be selected using information criteria, for example. When $\lambda$ is large enough, all coefficients are forced to be exactly zero. Inversely, $\lambda=0$ corresponds to the unpenalized ordinary least squares (OLS) estimate. When $p > n$ (the number of covariates is greater than the sample size), Lasso can select only $n$ covariates (even when more are associated with the outcome), and it tends to select one covariate from any set of highly correlated covariates.

Another popular penalty option can derived from $\Psi_p(\bbeta)=||\bbeta||^2_2=\sumasp\beta_j^2$ ($L_2$ norm), known as ridge  penalty \citep{Fu1998,Hoerl2000}. Ridge regression decreases the complexity of a model but does not reduce the number of variables since it never forces the coefficient to be zero but rather only minimizes it. Hence, this model is not good for feature reduction like the Lasso-regularized linear regression model. On the other hand, ridge regression tends to perform better than Lasso in scenarios with strongly correlated covariates, even when $p > n$.

Finally, \cite{zou2005regularization} introduced the elastic net, which uses the penalties from both the Lasso and ridge techniques to regularize regression models. The technique combines both the Lasso and ridge regression methods by learning from their shortcomings to improve the regularization of statistical models. The elastic net penalty is defined by
$$\Psi^{\alpha}_p(\bbeta)=\alpha||\bbeta||_1+(1-\alpha)||\bbeta||^2_2=\alpha\sumasp\beta_j^2+(1-\alpha)\sumasp|\beta_j|.$$
Elastic net is the same as Lasso when $\alpha = 1$. As $\alpha$ shrinks toward 0, elastic net approaches ridge regression. For other values of $\alpha$, the penalty term $\Psi_p^ {\alpha}(\bbeta)$ interpolates between the $L_1$ norm of $\beta$ and the squared $L_2$ norm of $\beta$. 

In \textsf{R}, an efficient implementation of regularization problems is available in the package \textbf{glmnet} \citep{friedman2010regularization}, which fits a generalized linear model via penalized ML. The regularization path is computed for Lasso or elastic net penalties at a grid of values for the regularization parameter. The algorithm is fast and exploits sparsity in the input matrix $\X$ to fit linear, logistic, multinomial, Poisson, and Cox regression models. A variety of predictions can be made from the fitted models. 

Our proposal takes advantage of this available software through the implementation of an EM algorithm for penalized estimation, which is discussed next.

\subsection{The EM algorithm for PML}\label{sec:EM}

The EM algorithm \citep{Dempster:1977} is a popular iterative algorithm for
ML estimation of models with incomplete data and has several appealing
features, such as stability of monotone convergence and simplicity of
implementation. Next, we discuss how the EM algorithm can be used for PML estimates of model parameters.

For general linear models, statistical inferences are based on underlying likelihood functions. The PML estimator can be used to select significant variables. Assume that the data $\{\Y_i,\X_i, \ii\}$ are collected independently and $\btheta=(\bbeta^{\top},\brho^{\top})^{\top}$ are the parameters of the model. Let $\ell(\btheta)=\ell(\bbeta,\brho)=\sumas \ell_i(\bbeta,\brho)$ denote the log-likelihood of $\btheta$ given observations $(\y_1,\x_1), \ldots,(\y_n,\x_n)$, a form of the penalized log-likelihood is
\begin{equation}\label{penlik}
\ell_p(\btheta)=\ell_p(\bbeta,\brho)=\ell(\bbeta,\brho)-\lambda\, \Psi_p(\bbeta).
\end{equation}

When the effect of a covariate is not signiﬁcant, the corresponding OLS estimate is often close but not equal to 0. Thus, this covariate is not excluded from the model. To avoid this problem, we may study submodels with various components of the design matrix excluded, as is done by forward and backward stepwise regression. However, the computational burden of these approaches is heavy and should be avoided. By maximizing (\ref{penlik}) that contains a penalty, there is a positive chance of having some estimated values of $\bbeta$ equaling 0 and thus of automatically selecting a submodel. Thus the procedure combines the variable selection and parameter estimation into one step and reduces the computational burden substantially.  

It is possible to apply the EM algorithm for penalized ML estimation \citep{green1990use} by assuming that unobserved components $\yp_L = (\yp^{\top}_{L_1},\ldots,\yp^{\top}_{Ln})^{\top}$ are hypothetical missing variables, and augmenting with the observed variables $\yp_\textrm{obs} = (\yp_{1}^{\top},\ldots,\yp_{n}^{\top})$. Hence, the penalized log-likelihood function for the model based on complete data $\yp_{c}=(\yp^{\top}_\textrm{obs}, \yp^{\top}_L)^{\top}$ is given by 
\begin{equation} \label{eq:loglikpcomp}
\ell^p_{c}(\bbeta,\brho|\yp_{c}) = \ell_{c}(\bbeta,\brho|\yp_{c}) - \lambda\,\Psi_p(\bbeta).
\end{equation}
 The EM algorithm maximizes (\ref{eq:loglikpcomp}) iteratively in the following
two steps:
\begin{itemize}
    \item[$\bullet$]{\bf E-step}. The E-step computes the conditional expectation of the function $\ell^p_{c}(\bbeta,\brho|\yp_{c})$ with respect to $\yp_{L}$, given the data $\yp_\textrm{obs}$ and assuming that the current estimate $\btheta^{(k)}=(\bbeta^{(k)},\brho^{(k)})$ gives  the true parameters of the model. The penalized $Q^p$-function is
    
    $$Q^p(\btheta|\btheta^{(k)})=\textrm{E}_{\scriptsize \btheta^{(k)}}[\ell^p_{c}(\btheta|\y_\textrm{obs},\y_L)|\y_\textrm{obs}]-\lambda\, \Psi_p(\bbeta),$$
     where  $\textrm{E}_{\scriptsize \btheta^{(k)}}$ means that the expectation is evaluated at for $\btheta=\btheta^{(k)}$ and  the superscript $(k)$ indicates the estimate of the related parameter  at the stage $k$ of the algorithm.
 In many applications, the $Q$-function can be written as
 \begin{eqnarray}
 \label{eqEM}
  Q^p(\btheta|\btheta^{(k)})&=&f(\brho,\btheta^{(k)})Q_1^{p}(\bbeta,\lambda|\btheta^{(k)})+Q^p_2(\brho|\btheta^{(k)}),
 \end{eqnarray}

 where $f(.)$ is a measurable real-valued function depending just on the parameter vector $\brho$, so that maximizing $Q^p(\btheta|\btheta^{(k)})$ with respect to $\bbeta$ is equivalent to maximizing $Q_1^{p}(\bbeta,\lambda|\btheta^{(k)})$ for $\lambda$ fixed, this is the kind of applications that we are interested here.
 
   \item[$\bullet$]{\bf M-step}.  The M-step on the $(k+1)$th iteration maximizes the function $Q^p(\btheta|\btheta^{(k)})$ with respect to $\theta$. From (\ref{eqEM}), keeping $\lambda$ fixed, the parameters are updated by
\begin{eqnarray*}
{\bbeta}^{(k+1)}(\lambda)&=&\underset{\bbeta}{\arg\min}\left\{Q^p_1(\bbeta,\lambda|\btheta^{(k)})\right\},\\
{\brho}^{(k+1)}&=&\underset{\brho}{\arg\min}\left\{Q^p(\bbeta^{(k+1)},\brho|\btheta^{(k)})\right\}.
\end{eqnarray*}
The algorithm is terminated when the relative distance between two successive evaluations of the penalized log-likelihood defined in (\ref{penlik}) is less than a tolerance value, such as {\small $|\ell_p({\btheta}^{(k+1)})/\ell_p({\btheta}^{(k)})-1|<\epsilon$}, with {\small $\epsilon=10^ {-6}$}. 
\end{itemize}

Although $\lambda$ is fixed for these steps, note that we can select its optimal value based on information criteria, for example. A more detailed discussion for selecting the tuning parameters in an LMM context is given in Section \ref{sec:lambda}.

\section{The linear mixed-effects model}\label{sec:model}

The classical normal LMM is specified as follows\citep{Laird:1982}:
\begin{equation}
\Y_i=\textbf{X}_i\bbeta+\textbf{Z}_i\textbf{b}_i+\bepsilon_i,
\label{modeleq}
\end{equation}
where $\textbf{b}_i\buildrel iid\over\sim
N_q(\mathbf{0},\textbf{D})$ is independent of $\bepsilon_i\buildrel
ind.\over\sim
N_{n_i}(\mathbf{0},\sigma^2\mathbf{I}_{n_i}),\,\ii;$  the
subscript $i$ is the subject index; 
$\textbf{Y}_i=(Y_{i1},\ldots,Y_{in_i})^{\top}$ is a $n_i\times 1$
vector of observed continuous responses for subject $i$;
$\textbf{X}_i$ is the $n_i\times p$ design matrix corresponding to
the fixed effects, $\bbeta$, of dimension $p\times 1$;
$\textbf{Z}_i$ is the $n_i\times q$ design matrix corresponding to
the $q\times 1$ vector of random effects $\textbf{b}_i$;
$\bepsilon_i$ of dimension $(n_i\times 1)$ is the vector of random
errors; and the dispersion matrix $\textbf{D}=\textbf{D}(\balpha)$
depends on unknown and reduced parameters $\balpha$. 

To account for high-dimension problems, we allow the general framework where the number $p$ of fixed-effects regression coefficients can be larger than the total number of observations, that is, $n>p$. 
To perform PML estimation in the general LMM specification from \eqref{modeleq}, we now present a proposal based on the EM algorithm.
\begin{algorithm}[ht!]
\begin{algorithmic}
\State For a fixed value of $\lambda$:
\State Initialization
\State $ {\bbeta}^{(0)}\leftarrow \underset{\bbeta}{\arg\min}\left\{(\Y-\X\bbeta)^{\top}(\Y-\X\bbeta)+\lambda ||\bbeta||_1\right\} $\\
	$\sigma^{2(0)}\leftarrow \frac{1}{n}(\Y-\X{\bbeta}^{(0)})^{\top}(\Y-\X{\bbeta}^{(0)})$,\,\, $\bD^{(0)}\leftarrow \mathbf{I}_ q$,\,\,$\textbf{b}^{(0)}\leftarrow \mathbf{0}$.\\
{\bf While} {the stopping criterion is not satisfied} {\bf do}
\State \\{\underline{E-Step:}\\
$  \tilde{\yp}^{(k)}_i= \yp_i-\Z_i{\ba}^{(k)}_i,\,\,\, {\bLambda}^{(k)}_i=({\bD}^{-1(k)}+\Z_i^{\top}\Z_i/{\sigma^{2}}^{(k)})^{-1},$\\
$  {\ba_i}^{(k)}=\frac{1}{{\sigma^2}^{(k)}}{\bLambda}^{(k)}_i\Z^{\top}_i(\tilde{\yp}^{(k)}_i-\X_i{\bbeta}^{(k)}),$\\ $\,\,\lambda^{(k)}_1=2\lambda \sigma^{2(k)}$.\\\vskip.3cm
\underline{M-Step:}\\
	  $ {\bbeta}^{(k+1)}\leftarrow \underset{\bbeta}{\arg\min}\left\{(\tilde{\yp}^{(k)}-\X\bbeta)^{\top}(\tilde{\yp}^{(k)}-\X\bbeta)+\lambda^{(k)}_1 \Psi_p(\bbeta)\right\}, $\\
	$\sigma^{2(k+1)}\leftarrow \frac{1}{N}\left[(\tilde{\yp}^{(k)}-\X\bbeta^{(k)})^{\top}(\tilde{\yp}^{(k)}-\X\bbeta^{(k)})+\sumas\Tr(\Z_i{\bLambda}^{(k)}_i\Z_i^{\top})\right],$\\
$
{\bD}^{(k+1)}\leftarrow \frac{1}{n}\sumas
({\ba}^{(k)}_i{{\ba}^{\top(k)}_i}+{\bLambda}^{(k)}_i).$
	}\\
{\bf end}

\State	Use BIC to select the optimal value of $\lambda$ on a grid of possible values \\
\Return $\btheta=(\bbeta,\sigma^2,\bD)$
\end{algorithmic}
 

	\caption{EMLMLasso}\label{cap1:algo::mean}
\end{algorithm}

\subsection{PML estimation in LMM} 
Let $\Y=(\Y^{\top}_1,\ldots,\Y^{\top}_n)^{\top}$ and
$\mathbf{b}=(\mathbf{b}^{\top}_1,\ldots,\mathbf{b}^{\top}_n)^{\top}$.  In the estimation procedure, $\mathbf{b}$ are treated as hypothetical missing data and augmented with the
observed data set, we have the complete data $\y_c=(\y^{\top},\mathbf{b}^{\top})$.
The EM-type algorithm is applied to the complete data penalized 
log-likelihood function 
\begin{eqnarray}
\ell^p(\btheta|\yp_c)&=&c-\frac{1}{2}\left[N\log{\sigma^2}+\frac{1}{\sigma^2}\sumas(\yp_i-\X_i\bbeta-\Z_i\ba_i)^{\top}(\yp_i-\X_i\bbeta-\Z_i\ba_i)\right.\nonumber\\
&&\left.+ n\log{|\bD|}+\sumas\ba_i^{\top}\bD^{-1}\ba_i\right]-\lambda \Psi_p(\bbeta),
\end{eqnarray}
with $c$ being a constant  independent of the parameter vector $\btheta$ and $N=\sumas n_i$. Given the current estimate $\btheta={\btheta}^{(k)}$, the E-step calculates the conditional expectation of the complete  data log-likelihood function, which, apart from constants that do not depend on $\btheta$, is given by
\begin{eqnarray*}
Q^p(\btheta|{\btheta}^{(k)})&=&-\frac{N}{2}\log{\sigma^2}-\frac{1}{2\sigma^2}\sumas\left[(\yp_i-\X_i{\bbeta}-\Z_i{\ba}^{(k)}_i)^{\top}(\yp_i-\X_i{\bbeta}-\Z_i{\ba}^{(k)}_i)\right.\\&&\left. + \Tr(\Z_i{\bLambda}^{(k)}_i\Z_i^{\top})\right]-\lambda \Psi_p(\bbeta)-\frac{n}{2}\log{|\bD|}-\frac{1}{2}\sumas\Tr\left(({\ba}^{(k)}_i{{\ba}^{\top(k)}_i}+{\bLambda}^{(k)}_i)\nonumber\bD^{-1}\right)\\
&=&- \frac{1}{2\sigma^2}\left((\tilde{\yp}^{(k)}-\X{\bbeta})^{\top}(\tilde{\yp}^{(k)}-\X{\bbeta})+\lambda_1\Psi_p(\bbeta)\right)\\
&&-\frac{N}{2}\log{\sigma^2}-\frac{1}{2\sigma^2}\sumas\Tr(\Z_i{\bLambda}^{(k)}_i\Z_i^{\top})-\frac{n}{2}\log{|\bD|}-\frac{1}{2}\sumas\Tr\left(({\ba}^{(k)}_i{{\ba}^{\top(k)}_i}+{\bLambda}^{(k)}_i)\nonumber\bD^{-1}\right) \\
&=&f(\sigma^2,{\btheta}^{(k)}) {Q^p_{1}(\bbeta, \lambda_1|{\btheta}^{(k)})}+{{
{Q^p_{2}(\balpha,\sigma^2|{\btheta}^{(k)})}}},
\end{eqnarray*}
where $f(\sigma^2,{\btheta}^{(k)})=\displaystyle\frac{1}{2\sigma^2}$, ${Q^p_{1}(\bbeta, \lambda_1|{\btheta}^{(k)})}=(\tilde{\yp}^{(k)}-\X{\bbeta})^{\top}(\tilde{\yp}^{(k)}-\X{\bbeta})+\lambda_1\Psi_p(\bbeta)$ and
$$
{{Q^p_{2}(\balpha|{\btheta}^{(k)})}}=-\frac{N}{2}\log{\sigma^2}-\frac{1}{2\sigma^2}\sumas\Tr(\Z_i{\bLambda}^{(k)}_i\Z_i^{\top}) -\frac{n}{2}\log{|\bD|}-\frac{1}{2}\sumas\Tr\left(({\ba}^{(k)}_i{{\ba}^{\top(k)}_i}+{\bLambda}^{(k)}_i)\nonumber\bD^{-1}\right),
$$
with $\X$ being the $N\times p$ matrix, $\X=(\X_1,\ldots \X_n)^{\top}$, 
$\tilde{\yp}^{(k)} =(\tilde{\yp}^{(k)} _1,\ldots,\tilde{\yp}^{(k)} _n)^{\top}$ the $N\times 1$ vector with elements $\tilde{\yp}^{(k)}_i= \yp_i-\Z_i{\ba}^{(k)}_i$, 
${\ba_i}^{(k)}=\textrm{E}_{\scriptsize \btheta^{(k)}}[\displaystyle\ba_i|\y_i]=\frac{1}{{\sigma^2}^{(k)}}{\bLambda}^{(k)}_i\Z^{\top}_i({\yp}_i-\X_i{\bbeta}^{(k)})$,
${\bLambda}^{(k)}_i=\textrm{Cov}[\ba_i|\y_i,{\btheta}^{(k)}]=({\bD}^{-1(k)}+\Z_i^{\top}\Z_i/{\sigma^{2}}^{(k)})^{-1}$, and $\lambda^{(k)}_1=2\lambda \sigma^{2(k)}$.

The M-step then conditionally maximizes
$Q(\btheta|{\btheta}^{(k)})$ with respect to $\btheta$ and
obtains a new estimate ${\btheta}^{(k+1)}$, as follows:
\begin{eqnarray}
 \beta^{(k+1)}(\lambda)&=&\underset{\bbeta}{\arg\min}\left\{(\tilde{\yp}^{(k)}-\X\bbeta)^{\top}(\tilde{\yp}^{(k)}-\X\bbeta)+\lambda^{(k)}_1 \Psi_p(\bbeta)\right\},   \\
{\sigma^2}^{(k+1)}&=&\frac{1}{N}\left[(\tilde{\yp}^{(k)}-\X\bbeta^{(k)})^{\top}(\tilde{\yp}^{(k)}-\X\bbeta^{(k)})+\sumas\Tr(\Z_i{\bLambda}^{(k)}_i\Z_i^{\top})\right],\,\, \\
{\bD}^{(k+1)}&=&\frac{1}{n}\sumas
({\ba}^{(k)}_i{{\ba}^{\top(k)}_i}+{\bLambda}^{(k)}_i).
\end{eqnarray}
Note that the M-step update of $\bbeta$ at each iteration is equivalent to the penalized estimator given in (\ref{PLSequation}) for the general linear regression model defined in \ref{ML1}, for which a super efficient and reliable algorithm is available in the \textsf{R} package \textbf{glmnet}.  Algorithm \ref{cap1:algo::mean} specifies the procedure, where the optimization with respect to $\bbeta$ is done through the \textsf{R} package \textbf{glmnet}.

\subsection{Choice of the tuning parameters}\label{sec:lambda}
In order to choose the optimal value of the tuning parameter, we use BIC \citep{wang2007tuning}. Thus, for a given tuning parameter $\lambda$, let $\hat\btheta=(\hat\bbeta,\hat\brho)$ be the penalized ML estimator obtained via the EM algorithm. The optimal set of $\btheta$ is selected by minimizing the following criterion:
$$BIC(\lambda)=-2\ell(\hat \bbeta,\hat\brho)+\log{(n)}\ {\widehat{df}}_{\lambda},$$
where  $\widehat{df}_{\lambda}$ is the number of nonzero elements of $\hat\bbeta$ plus the number of parameters on $\hat \brho$, and $n$ is the number of subjects. The Akaike information criterion (AIC) is another popular criterion used in variable selection and model selection and is well known to select the model with the optimal prediction performance, while BIC is generally preferred when aiming to select the true sparse model \citep{zou2007degrees}. 

\section{Simulation studies} \label{sec:experiments}

In this section, we examine the performance of the proposed procedure, denoted as the \texttt{EMLMLasso} algorithm under three scenarios, and compare the simulation results with the \texttt{glmmLasso} algorithm \citep{Groll2022}. Following \cite{PanShang2018}, we generated 100 datasets from the model (\ref{modeleq}) for each scenario. The results were obtained using the \textsf{R} software, and the codes are available on GitHub.

\subsection{Scenario 1}

We consider the true model with $p = 9$ for fixed effects, $q = 2$ for random effects, and the true value of the parameters are set at $\bbeta = (1,1,0,0,0,0,0,0,0)^{\top}$ for the fixed effects, and

\begin{equation}\label{Dsimulate}
	\bD = \begin{pmatrix}
		1.0   & 0.25 \\
		0.25  & 1.0
	\end{pmatrix},
\end{equation}
 for the variance-covariance matrix. The first column of $\mathbf{Z}_i$ consists of ${1}$'s for the subject-specific intercept, and the second column is a sequence of discrete values from $1$ to $n_i$. The columns of the matrix $\mathbf{X}_i$ are independently generated from a normal distribution with mean equals $6$ and variance equals $1$. Additionally, both algorithms' column values of the matrix $\mathbf{X}_i$ are centered. We further assume the variance for the residuals $\sigma^2=1$. First, we use $n = 30$ and $n_i = 5$. Then, we examine the performance of the proposed procedure in a larger sample scenario, considering $n=60$ and $n_i = 10$.

For each $\beta_j, j = 1, 2, \dots, p$, we calculate the proportion of times that $\beta_j$ estimates are $0$ using both \texttt{EMLMLasso} and \texttt{glmmLasso}, with the BIC criterion to select the optimal value of the tuning parameter $\lambda$. For the optimal value $\lambda$ in the \texttt{glmmLasso}, we kept the same sequence suggested by \cite{Groll2022}, a sequence from 500 to 0 by $-5$, while for the \texttt{EMLMLasso} we consider a sequence from 0.001 to 0.5 with length out equal to $100$. 

In addition, we compute the root mean squared error (RMSE) in each sample to inspect the performance of the proposed method. This estimation measure quantifies the difference between fixed effects parameters and their estimates. According to \cite{Lee2022} the RMSE is 
\begin{equation}\label{RMSE}
\text{RMSE} = \sqrt{(\hat{\bbeta} - \bbeta)^{\top} (\hat{\bbeta} - \bbeta)/M},
\end{equation}
where $M$ is the number of Monte Carlo samples used in the simulation. The results for Scenario 1 are in Table \ref{Simulation12}.

\begin{center}
    \begin{table*}[htb]
      \caption{Simulation results (Scenario 1). The proportion of the number of times the zero coefficients obtained for 100 simulations and different sample sizes. RMSE was defined in \ref{RMSE}.}
       \label{Simulation12}
       \centering
      \begin{tabular*}{450pt}{@{\extracolsep\fill}ccccc@{\extracolsep\fill}}
\toprule
&\multicolumn{2}{c}{$n=30$, $n_i=5$}&\multicolumn{2}{c}{$n=60$, $n_i=10$}\\
		Parameter  &  \texttt{EMLMLasso} & \texttt{glmmLasso} &  \texttt{EMLMLasso} & \texttt{glmmLasso}\\
\midrule
			$\beta_{1}$ &   0    &    0       & 0     &     0\\
                $\beta_{2}$ &   0    &    0       & 0     &     0\\
			$\beta_{3}$ &   0.88 &    0.24    & 0.92  &     0.33\\
			$\beta_{4}$ &   0.91 &    0.33    & 0.98  &     0.31\\
			$\beta_{5}$ &   0.90 &    0.29    & 0.98  &     0.30\\
                $\beta_{6}$ &   0.88 &    0.23    & 0.96  &     0.44\\
                $\beta_{7}$ &   0.94 &    0.29    & 0.96  &     0.31\\
			$\beta_{8}$ &   0.88 &    0.30    & 0.94  &     0.35\\
			$\beta_{9}$ &   0.87 &    0.39    & 0.89  &     0.25\\
\midrule
                RMSE        &   0.25 &    0.27    & 0.12  &     0.12\\
\bottomrule
      \end{tabular*}
\end{table*}
\end{center}

Table \ref{Simulation12} shows Scenario 1, where the proportion of the number of times the zero coefficients obtained for 100 simulations was recorded. Since the true fixed effects vector $\mathbf{\bbeta} = (1,1,0,0,0,0,0,0,0)^T$, we expect the results in the lines corresponding to $\beta_1$ and $\beta_2$ approximately equal to $0$ and in the other lines corresponding to $\beta_3$ to $\beta_9$ approximately equal to $1$.  Note from this table that both methods always correctly identified the significant variables ($\beta_1$ and $\beta_2$), while \texttt{glmmLasso} misclassifies them as 0 more than half of the times for both sample sizes. The RMSE results of \texttt{EMLMLasso} are less than or equal to \texttt{glmmLasso}, and both are better as the sample size increases because the smaller the value, the closer $\hat{\bbeta}$ is to the true parameter ${\bbeta}$.

\subsection{Scenario 2} 

Here, we aim to study the effect of categorical variables. In order to do that, the first column of $\mathbf{X}_i$ is generated from a Bernoulli distribution with $p=0.5$, and the other columns are generated independently from a normal distribution with mean equals $6$ and variance equals $1$. The column values of the matrix $\mathbf{X}_i$ corresponding to the normal distribution were standardized to have mean 0 and variance 1. Like the previous scenario, we examine the performance of the proposed procedure with categorical variables in a smaller sample ($n=30$ and $n_i = 5$), and after, we increase the sample size to $n=60$ and $n_i = 10$. We also used the same sequence presented the Scenario 1 for the $\lambda$ in both algorithms. The results are shown in Table \ref{Simulation78}.

\begin{table*}[htb]
      \caption{Simulation results (Scenario 2). The proportion of the number of times the zero coefficients obtained for 100 simulations and different sample sizes. RMSE was defined in \ref{RMSE}.}
       \label{Simulation78}
       \centering
        \begin{tabular*}{450pt}{@{\extracolsep\fill}ccccc@{\extracolsep\fill}}
\toprule
&\multicolumn{2}{c}{$n=30$, $n_i=5$}&\multicolumn{2}{c}{$n=60$, $n_i=10$}\\
		Parameter  &  \texttt{EMLMLasso} & \texttt{glmmLasso} &  \texttt{EMLMLasso} & \texttt{glmmLasso}\\
\midrule
			$\beta_{1}$ &   0    &    0       &  0     &    0\\
                $\beta_{2}$ &   0    &    0       &  0     &    0\\
			$\beta_{3}$ &   0.90 &    0.15    &  0.96  &    0.22\\
			$\beta_{4}$ &   0.91 &    0.19    &  0.97  &    0.15\\
			$\beta_{5}$ &   0.88 &    0.17    &  0.97  &    0.15\\
                $\beta_{6}$ &   0.94 &    0.15    &  0.98  &    0.20\\
                $\beta_{7}$ &   0.87 &    0.06    &  0.94  &    0.18\\
			$\beta_{8}$ &   0.91 &    0.14    &  0.93  &    0.13\\
			$\beta_{9}$ &   0.93 &    0.21    &  0.94  &    0.12\\
   \midrule
                RMSE        &   0.30 &    0.32    &  0.13  &    0.14\\
\bottomrule
		\end{tabular*}
\end{table*}

For Scenario 2, as the true fixed effects vector $\bbeta = (1,1,0,0,0,0,0,0,0)^T$, again we expect that the proportions in the rows corresponding to $\beta_1$ and $\beta_2$ are close to $0$ and from $\beta_3$ to $\beta_9$ close to $1$. From Table \ref{Simulation78}, we can see that even with categorical variables in the model, both algorithms correctly identified the significant variables ($\beta_1$ and $\beta_2$), and the proposed algorithm excelled in excluding irrelevant variables. The RMSEs are smaller in \texttt{EMLMLasso}, indicating that the proposed algorithm outperforms \texttt{glmmLasso}.

\subsection{Scenario 3} 

 The proposed approach can also handle high-dimensional predictors. Thus, in this setting,  we evaluate the effect of increasing the dimension of $\bbeta$ for a vector $p \times 1$, with $p = 50$. The $\bbeta$ are set as $\bbeta = (1, ..., 1, 0, ...,0)^{\top}$, where the first ${p}^*$ elements of $\bbeta$ are equal to 1 and the remaining $(p-{p}^*)$ are equal to 0. We evaluated ${p}^*= 5$ and ${p}^*=10$. 
 
 The columns of the matrix $\mathbf{X}_i$ are independently generated from a normal distribution with mean equals $6$ and variance equals $1$. We considered $\sigma^2 = 1$, $n=30\,\,(<p)$ (high-dimensional data) and $n=60$, $n_i=5$ and $n_i=10$, and $\bD$ as in (\ref{Dsimulate}) with
\begin{equation}\label{Dsimulate2}
	\bD = \begin{pmatrix}
		9   & 4.8 \\
		4.8 & 4
	\end{pmatrix}.
\end{equation}
For the estimation of $\lambda$ in the \texttt{glmmLasso}, we kept the same sequence of the previous scenarios, i.e., a sequence from 500 to 0 by $-5$, and the \texttt{EMLMLasso} we considered a sequence from 0.001 to 0.5 with length out equal to $100$. 

As the performance measures of the variable selection, following \cite{ChunKeles2010}, we calculated the average sensitivity and specificity, defined by
\begin{itemize}\setlength{\itemsep}{0pt}
    \item[$\bullet$]{\textbf{sensitivity:}} the proportion of nonzero estimates among the true nonzero elements of $\bbeta$;
    \item[$\bullet$]{\textbf{specificity:}} the proportion of zero estimates among the true zero elements of $\bbeta$.
\end{itemize}
Perfect variable selection occurs when both the sensitivity and specificity are equal to one, and in a good variable selection method, both measures need to be large. 

Let ${p}^*$ be the number of nonzero elements of true $\bbeta$. The sensitivity was calculated as the average number of nonzero $\bbeta$ estimates divided by $p^*$. The specificity was calculated as the average number of zero $\beta$ estimates divided by $({p} - {p}^*)$, with $p=50$. The results for ${p}^* = 5$ and $p^* = 10$ are given in Table \ref{SimulationTab678}.

\begin{table*}[htb]
\caption{Simulation results for Scenario 3 when $p=50$. The numbers in parentheses indicate where the $\bD$ matrices have been defined, and the RMSE has been defined in \ref{RMSE}.}
       \label{SimulationTab678}
       \centering
       \begin{tabular*}{450pt}{@{\extracolsep\fill}cccccrrr@{\extracolsep\fill}}
\toprule
			Sample size           & $n_i$                & $p^*$       & $\bD$  
                & Algorithm           & Sensitivity          & Specificity          &  RMSE\\
\midrule			\tabularnewline
			\multirow{10}{*}{30}  & \multirow{10}{*}{5}  & \multirow{4}{*}{5}   & \multirow{2}{*}{(\ref{Dsimulate})}
                 & \texttt{EMLMLasso}       & 1.00                 & 0.92                 & 0.52
                 \tabularnewline
				                   &                      &                      &  
                & \texttt{glmmLasso}         & 0.92                 & 0.50                 & 0.93\\
                 \cline{4-8}
                 \tabularnewline
	                               &                      &                      & \multirow{2}{*}{(\ref{Dsimulate2})}
                & \texttt{EMLMLasso}        & 1.00                 & 0.92                 & 0.55
                \tabularnewline
				                   &                      &                      &  
                & \texttt{glmmLasso}         & 0.39                 & 0.73                 & 1.83\\
                 \cline{3-8}
                 \tabularnewline
                                      &                      & \multirow{4}{*}{10}   & \multirow{2}{*}{(\ref{Dsimulate})}
                & \texttt{EMLMLasso}        &  1.00                &  0.72                & 0.55
                \tabularnewline
				                   &                      &                      &  
                & \texttt{glmmLasso}         &  1.00                &  0.05                & 0.88\\
                 \cline{4-8}
                 \tabularnewline
                                      &                      &                      & \multirow{2}{*}{(\ref{Dsimulate2})}
                & \texttt{EMLMLasso}        &  1.00                &  0.71                & 0.58
                \tabularnewline
				                   &                      &                      &  
                & \texttt{glmmLasso}         &  0.99                &  0.03                & 1.00\\
            \midrule
                 			\tabularnewline
			\multirow{10}{*}{60}  & \multirow{10}{*}{10} & \multirow{4}{*}{5}   & \multirow{2}{*}{(\ref{Dsimulate})} 
                 & \texttt{EMLMLasso}       & 1.00                 & 0.97                 & 0.23
                 \tabularnewline
				                   &                      &                      &  
                & \texttt{glmmLasso}         & 1.00                 & 0.34                 & 0.29\\
                 \cline{4-8}
                 \tabularnewline
	                               &                      &                      & \multirow{2}{*}{(\ref{Dsimulate2})}
                & \texttt{EMLMLasso}        &  1.00                &  0.97                & 0.23
                \tabularnewline
				                   &                      &                      &  
                & \texttt{glmmLasso}         &  1.00                &  0.02                & 0.33\\
                 \cline{3-8}
                 \tabularnewline
                                      &                      & \multirow{4}{*}{10}  & \multirow{2}{*}{(\ref{Dsimulate})}
                & \texttt{EMLMLasso}        & 1.00                 & 0.93                 & 0.28
                \tabularnewline
				                   &                      &                      &  
                & \texttt{glmmLasso}         & 1.00                 & 0.05                 & 0.34\\
                 \cline{4-8}
                 \tabularnewline
                                      &                      &                      & \multirow{2}{*}{(\ref{Dsimulate2})}
                & \texttt{EMLMLasso}  & 1.00                 &  0.93                & 0.29
                \tabularnewline
				                   &                      &                      &  
                & \texttt{glmmLasso}         & 1.00          &  0.02                & 0.33\\
          \bottomrule
		\end{tabular*}
		\protect\protect
		{\small{}}
	\end{table*}

From Table \ref{SimulationTab678}, we can see that \texttt{EMLMLasso} always gets the significant variables correctly since the sensitivity is equal to 1 under different sample sizes. Among the zero $\bbeta$ estimates, \texttt{EMLMLasso} shows that it has better results for a large sample size. However, in \texttt{glmmLasso} there was a trade-off between the two measures. In general, the algorithm produced better sensitivity and worse specificity; only when $n=30$, $n_i=5$, and $\bD$ as in (\ref{Dsimulate2}), the algorithm showed better specificity and worse sensitivity.

In addition, we used 10-fold cross-validation to evaluate the performance of the algorithms when $p > n$. The 10-fold cross-validation technique is used in machine learning to evaluate some performance \citep{Stone1974,Wasserman2006}. We considered the second configuration of the Table \ref{SimulationTab678} to obtain the 100 Monte Carlo (MC) datasets, i.e., $n=30$, $n_i=5$, $p=50$, $\bbeta = (1,1,1,1,1,0, ...,0)^{\top}$, where the first $p^*=5$ elements of $\bbeta$ are equal to 1 and $(p-p^*)$ are equal to 0, $D$ as in (\ref{Dsimulate2}), and $\lambda$ according to the previous scenarios. The technique divides the datasets into 10 equal parts or folds, with one fold used for testing and the other nine for training the model. This process is repeated ten times, each time using a different fold for testing, and the average performance is calculated through RMSE. By using 10-fold cross-validation with RMSE as in (\ref{RMSE}), we evaluated through the boxplot which of the algorithms provided a smaller value for this quantity, implying that it is the algorithm with the best performance in the selection for fixed effects in linear mixed-effects models. The 10-fold cross-validation was implemented  by using the available \textsf{R} packages  \textbf{joineR}, \textbf{lme4}, \textbf{splines}, and \textbf{caret}.

\begin{figure}[htb]
\centering
\includegraphics[width=10cm]{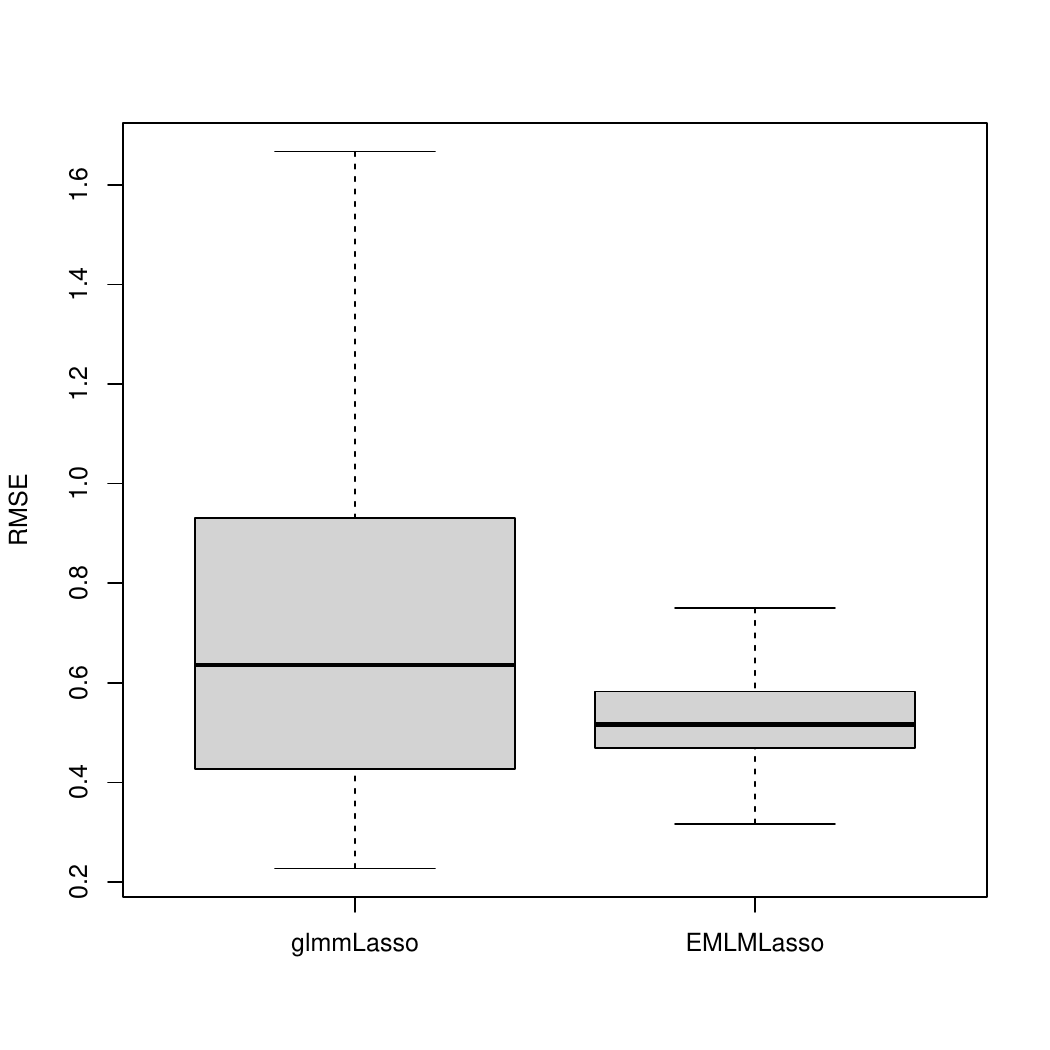}
\caption[FIGURE 2]{Simulation results for the first configuration of Scenario 3. 10-fold cross-validation to evaluate the performance of the algorithms when $p>n$, with $p=50$, $n=30$, $n_i=5$, $D$ as in (\ref{Dsimulate}).}
\label{CrossValid}
\end{figure}

From Figure \ref{CrossValid}, we see that the distribution of the RMSE is closer to zero for the \texttt{EMLMLasso}, indicating that the proposed method provides better predictive power.

\section{Case Studies}\label{sec:realdata}

We illustrate the proposed methods with the analysis of
two  datasets.

\subsection{Framingham cholesterol data}
For illustration, we applied the algorithms presented to the Framingham cholesterol data. The Framingham heart study has examined the role of serum cholesterol as a risk factor for the evolution of cardiovascular diseases. \cite{Zhang:2001} proposed a semiparametric approach to analyze a subset of the Framingham cholesterol data, which consists of sex, baseline age, and cholesterol levels measured at the beginning of the study and then every two years over a period of 10 years, for 200 randomly selected participants. After that, the data were analyzed by \cite{Arellano:2005}, \cite{Lachos:2007}, \cite{Lin:2008}, \cite{Lachos:2010}, among others. Here, we revisit this dataset with the aim of applying \texttt{EMLMLasso} and compare the results with \texttt{glmmLasso}.
Assuming a linear growth model with subject-specific random intercept and slopes, we fit an LMM model to the data: 
\begin{equation}\label{ap1}
\begin{split}
    y_{ij} =&  \beta_1 sex_i + \beta_2 age_i + \beta_3 t_{ij} +  \beta_4 sex_i\times age_i + \beta_5 sex_i\times t_{ij} +\beta_6 age_i\times t_{ij} + \\ 
    &+\beta_7 sex_i\times age_i\times t_{ij} +\beta_8 x_{1j} + \beta_9 x_{2j}+ \beta_{10} x_{3j} 
    + b_{0i} + b_{1i} t_{ij} + \epsilon_{ij},
\end{split}
\end{equation}
where $y_{ij}$ is the cholesterol level centered at its sample mean and divided by 100 at the $j$th time for subject $i$; $t_{ij}$ is $(time - 5)/10$, with time measured in years from the start of the study; $age_i$ is the age at the start of the study; and $sex_i$ is the sex indicator (0 = female, 1 = male), as described in Table \ref{DescAp1}.

\begin{table*}[htb]
      \caption{Framingham cholesterol data. Description of original covariates.}
       \label{DescAp1}
       \centering
      \begin{tabular*}{450pt}{@{\extracolsep\fill}ll@{\extracolsep\fill}}
\toprule
Variable  &  Description\\
\midrule
			Cholesterol $(y_{ij})$   & Cholesterol level,  centered at its mean and divided by 100 \\
                                         & (continuous, mean = 2.34 , standard deviation = 0.46, $1.29 \leq y_{ij} \leq 4.3$)\\
                Sex      $(sex_i)$    & Sex indicator for subject $i$, with 0 is female (51\%), and 1 is male (49\%)\\
			Age         $(age_i)$    & Age of the participant $i$ in years\\ 
                                         & (continuous, mean = 42.47, standard deviation = 7.89, $31 \leq age_i \leq 62$)\\
			Time        $(t_{ij})$   & Time measured in years from the start of the study, $(time - 5)/10$, with values: \\ &-0.5 (19.2\%), -0.3 (16.9\%), -0.1 (16.2\%),  0.1 (16.2\%),  0.3 (16.1\%), and  0.5 (15.5\%)\\ 
\bottomrule
		\end{tabular*}
\end{table*}

We considered in the model all interaction terms between sex, age, and time, and to evaluate the performance of each algorithm, we added 3 simulated covariates generated independently of the response variable: one categorical and two correlated normal variables. The first covariate was generated from a Bernoulli distribution with $p = 0.5$, and the second and third covariates were generated from a bivariate standard normal distribution with correlation $\rho=0.5$. We standardized each covariate, except for the sex and Bernoulli categorical covariates, and centered the response variable at the sample mean to drop the fixed effects intercept.

\begin{figure}[htb]
\centering
\includegraphics[width=13cm]{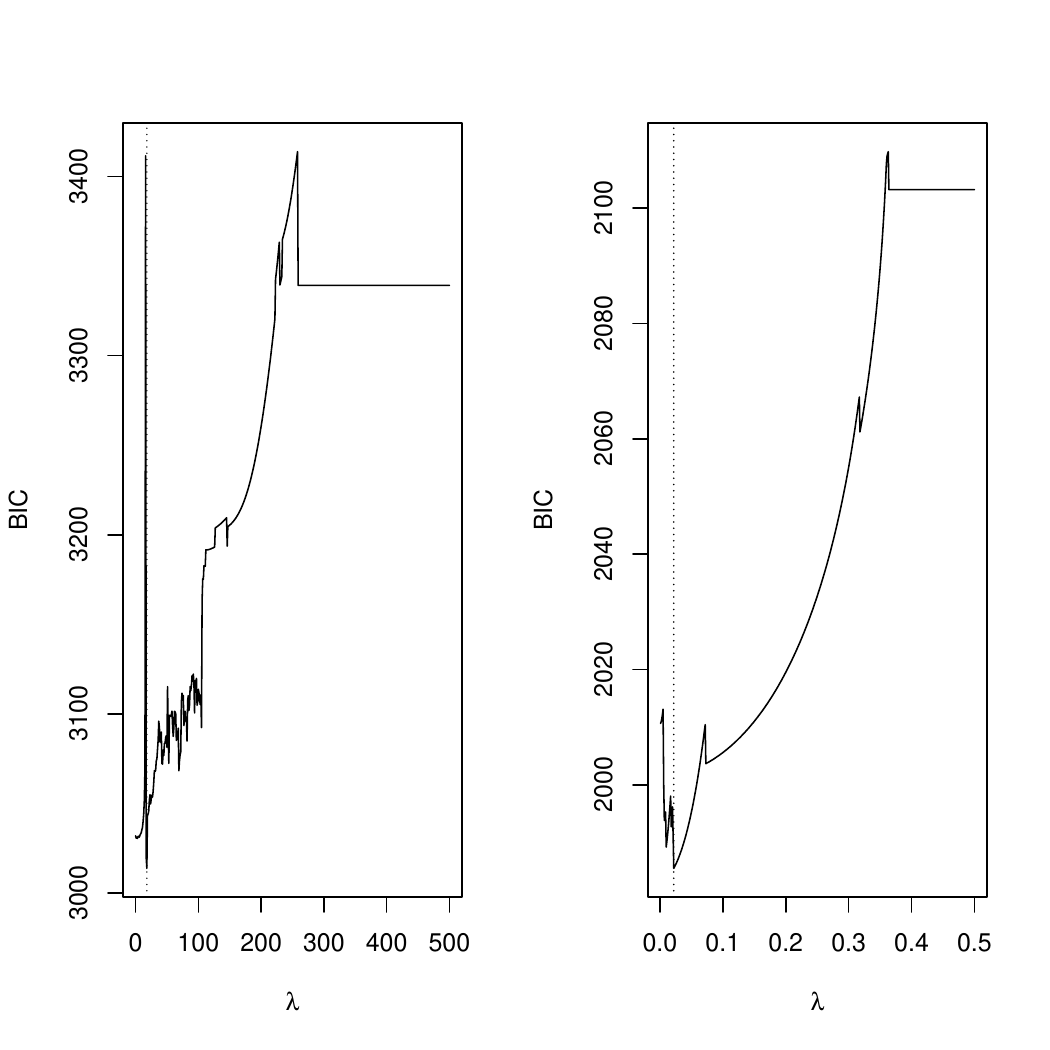}
\caption[FIGURE 3]{Framingham cholesterol data. Results for BIC for the \texttt{glmmLasso} (left panel) and \texttt{EMLMLasso} (right panel) as a function of the penalty parameter $\lambda$, considering the covariates given in Table \ref{Result1Ap1}.}
\label{BICallcov}
\end{figure}

\begin{table*}[htb] 
      \caption{Framingham cholesterol data. Selected variables with coefficient estimates by \texttt{EMLMLasso} and \texttt{glmmLasso}, and p-value by \textbf{lme4}.}
       \label{Result1Ap1}
       \centering
       \begin{tabular*}{450pt}{@{\extracolsep\fill}lrrrr@{\extracolsep\fill}}
\toprule
Variable                    &  \texttt{EMLMLasso} &  p-value (\textbf{lme4})  &  \texttt{glmmLasso}  &  p-value (\textbf{lme4})\\
\midrule
Intercept                   &    $-$              &   $<0.0001$     &    $-$               &  $<0.0001$ \\
Sex                         &   0.0000            &        $-$       &  -0.0737             &  0.0453\\
Age                         &   0.2219            &   $<0.0001$     &   0.0173             &  0.1992\\
Time                        &   0.1485            &   $<0.0001$     &  -0.2030             &  0.0022\\
Sex$*$Age                   &   0.0000            &        $-$       &   0.0442             &  0.0439\\
Sex$*$Time                  &   0.0813            &   $0.0004$     &   0.2733             &  0.0977\\
Age$*$Time                  &   0.0000            &        $-$       &   0.2038             &  0.0471\\
Sex$*$Age$*$Time            &   0.0000            &        $-$       &  -0.3092             &  0.3431\\
Bernoulli$(0.5)$            &   0.0000            &        $-$       &   0.0784             &  $-$\\
Bivariate $\text{Normal}_1$ &   0.0000            &        $-$       &  -0.0324             &  $-$\\
Bivariate $\text{Normal}_2$ &   0.0000            &        $-$       &  -0.0045             &  $-$\\
\bottomrule
      \end{tabular*}
\end{table*}

In Figure \ref{BICallcov}, we plot the BIC against the smoothing parameter $\lambda$. The optimal values of the $\lambda$ are shown by the vertical line, i.e., $0.022$ and $18$, for \texttt{EMLMLasso} and \texttt{glmmLasso}, respectively.
Table \ref{Result1Ap1} shows that \texttt{EMLMLasso} selected age, time, and the interaction between sex and time, while \texttt{glmmLasso} selected all covariates. After selecting the fixed effects, we refit the model with the \textsf{R} packages \textbf{lme4} and \textbf{lmerTest} using the selected variables from each algorithm, except the simulated covariates. For this analysis, we did not remove the intercept and kept the original variables (without standardization) for ease of interpretation. We notice that the selected variables in the \texttt{EMLMLasso} are all significant since the p-value are very small. However, \texttt{glmmLasso} were not compatible with the results provided by the \textsf{R} packages \textbf{lme4} and \textbf{lmerTest}. 

\subsection{Riboflavin data}

Gene expression experiments study how genes are turned on and off and how this controls what substances are made in a cell. This dataset concerns the response of riboflavin (vitamin B2) production of bacillus subtilis (b. subtilis), a single celled organism (bacterium) found in the human digestive tract. The final goal of researchers is to increase the riboflavin production rate of b. subtilis by editing relevant genes. To facilitate this goal, we used the riboflavinV100 dataset, which contains the genes that most strongly influence the rate of riboflavin production \citep{Schelldorfer2011}. The data is provided by DSM (Switzerland) and made publicly available in the supplemental materials of \cite{Buhlmann2014}. This  dataset was previously analyzed by  \cite{Schelldorfer2011}, \cite{Buhlmann2014}, \cite{Bradic2020}, \cite{AlabisoShang2022}, among others. We also use \texttt{glmmLasso} to select relevant covariates for this dataset and compare the results with the ones obtained via \texttt{EMLMLasso}.

Given the longitudinal character of the dataset, we consider the following linear mixed-effects model:
 \begin{equation}\label{ap2}
y_{ij} = \sum_{k=1}^{100} \beta_k x_{ijk}  + \beta_{101} t_{ij} + b_{0i} + b_{1i} t_{ij} + \epsilon_{ij},
\end{equation}
where the response variable is the log of the rate of riboflavin produced, and there are 100 covariates representing the log of the expression level of 100 genes and the covariate time. This dataset consists of $n = 28$ different strains (species subtypes) of b. subtilis measured between two and four times over the course of 96 hours ($n_i \in \{2, 3, 4\}$), totalizing 71 observations. We standardize the response and all covariates to have mean zero and variance one.

\begin{figure}[htb]
\centering
\includegraphics[width=13cm]{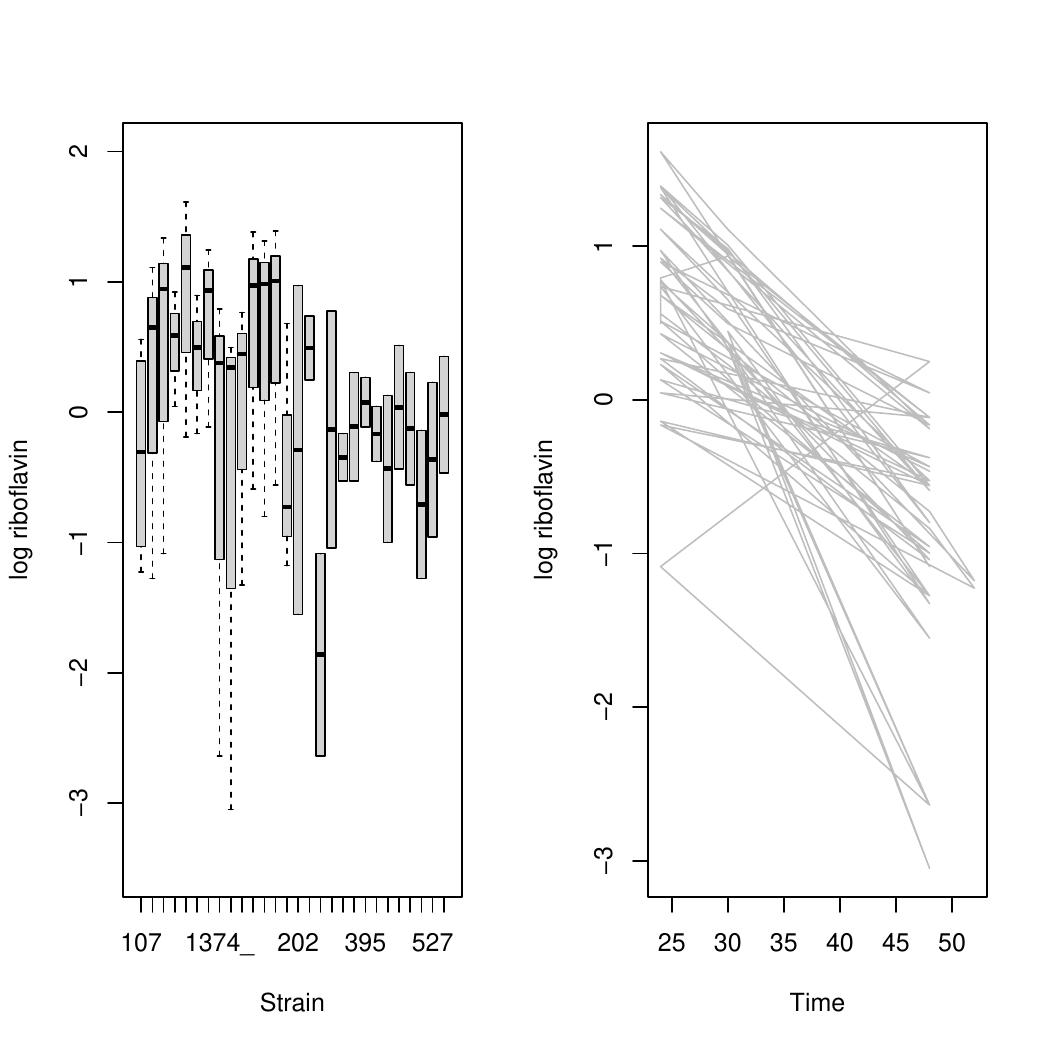}
\caption[FIGURE 4]{Riboflavin data. Boxplot of response variable over strains (left) and response over time (right).}
\label{Descript2}
\end{figure}

As in \cite{AlabisoShang2022}, the boxplot presented in Figure \ref{Descript2} (left panel) shows the response's distribution by the strains. We can observe that there is some difference in the response that appears to be attributable to the strain. Figure \ref{Descript2} (right panel) plots the response of the strains as a function of time. In general, we see that riboflavin decreases with time for each of these strains, indicating that time is likely to enter the model, as concluded by\cite{AlabisoShang2022}. We also notice from this figure that several strains whose response drops quickly over time, while others drop and then hold steady. Also, note that this dataset is unbalanced and not all strains are measured at the same points in time.

The first application to real data showed that even with correlated covariates, \texttt{EMLMLasso} presented satisfactory results. In this second application, the number of correlated covariates is larger. For this reason, we decided to evaluate the results of the algorithms with two methods: $1)$ complete matrix $\textbf{X}$, and $2)$ reduced matrix $\textbf{X}^*$. The matrix $\textbf{X}$, of dimension $71 \times 101$, is obtained from the riboflavinV100 dataset. The reduced matrix $\textbf{X}^*$, of dimension $71 \times 70$, is obtained using the package \texttt{findLinearCombos} in \textsf{R}, which removes columns that have  linear combinations among them in a matrix $\textbf{X}$.

For the estimation of $\lambda$ in the \texttt{glmmLasso}, we kept a sequence from 500 to 0, by -1, and the \texttt{EMLMLasso} we considered a sequence from 0.001 to 0.5 with length out equal to 500. When we work the complete matrix $\textbf{X}$ (Method 1), the optimal values are $43$ and $0.22$ for \texttt{glmmLasso} and \texttt{EMLMLasso}, respectively. For reduced matrix $\textbf{X}^*$ (Method 2), the optimal values are $43$ and $0.404$ for \texttt{glmmLasso} and \texttt{EMLMLasso}, respectively. Table \ref{DescAp2} shows that with Method 1, \texttt{EMLMLasso} selected 22 genes, and the \texttt{glmmLasso} selected 3 covariates. However, when we use Method 2, the \texttt{EMLMLasso} selected 12 genes and the covariate TIME, and the \texttt{glmmLasso} selected only 1 covariate.

\begin{table*}[htb]
{
\caption{Riboflavin data. Gene selections in riboflavinV100 dataset.}
       \label{DescAp2}
       \centering
         \begin{tabular*}{450pt}{@{\extracolsep\fill}ll@{\extracolsep\fill}}
\toprule
Methods  &  Variable list\\
\midrule
$^{(a)}$\texttt{glmmLasso}  & TIME  XHLA\_at  XHLB\_at \\
$^{(b)}$\texttt{findLinearCombos} + \texttt{glmmLasso}    &  TIME\\
$^{(c)}$\texttt{EMLMLasso}  &  YHZA\_at YHFH\_r\_at NADC\_at YPUF\_at ACOA\_at YPUD\_at \\
                    &  YCGN\_at YXLE\_at YTGD\_at PURC\_at XLYA\_at YCGO\_at \\
                    &  GSIB\_at YTCF\_at GAP\_at YRDD\_i\_at CARA\_at YCIB\_at \\
                    &  YOSJ\_at ALD\_at TRXA\_at PCKA\_at \\
$^{(d)}$\texttt{findLinearCombos} + \texttt{EMLMLasso}    &  TIME YHZA\_at YRZI\_r\_at DEGQ\_r\_at YXLE\_at ARGF\_at \\
                                                   & YTGD\_at GUAB\_at AHPC\_at XLYA\_at YCGO\_at YTCF\_at GAP\_at\\
\bottomrule
		\end{tabular*}
  }
\end{table*}

We use the \texttt{riboflavinV100} dataset, the \textsf{R} package \textbf{lme4}, and fitted a mixed-effect model with the predictors obtained with Method 1 and another with the predictors from Method 2, for each algorithm. We used the \textsf{R} packages \textbf{joineR}, \textbf{lme4}, \textbf{splines}, and \textbf{caret} to perform the 4-fold cross-validation and compare the predictive power of each method, calculating the mean squared error of $\textbf{y}$
\begin{eqnarray*}
    \text{MSE}_y = (\by - \hat{\by})^{\top} (\by - \hat{\by}).
\end{eqnarray*}

\begin{figure}[htb]
\centering
\includegraphics[width=10cm]{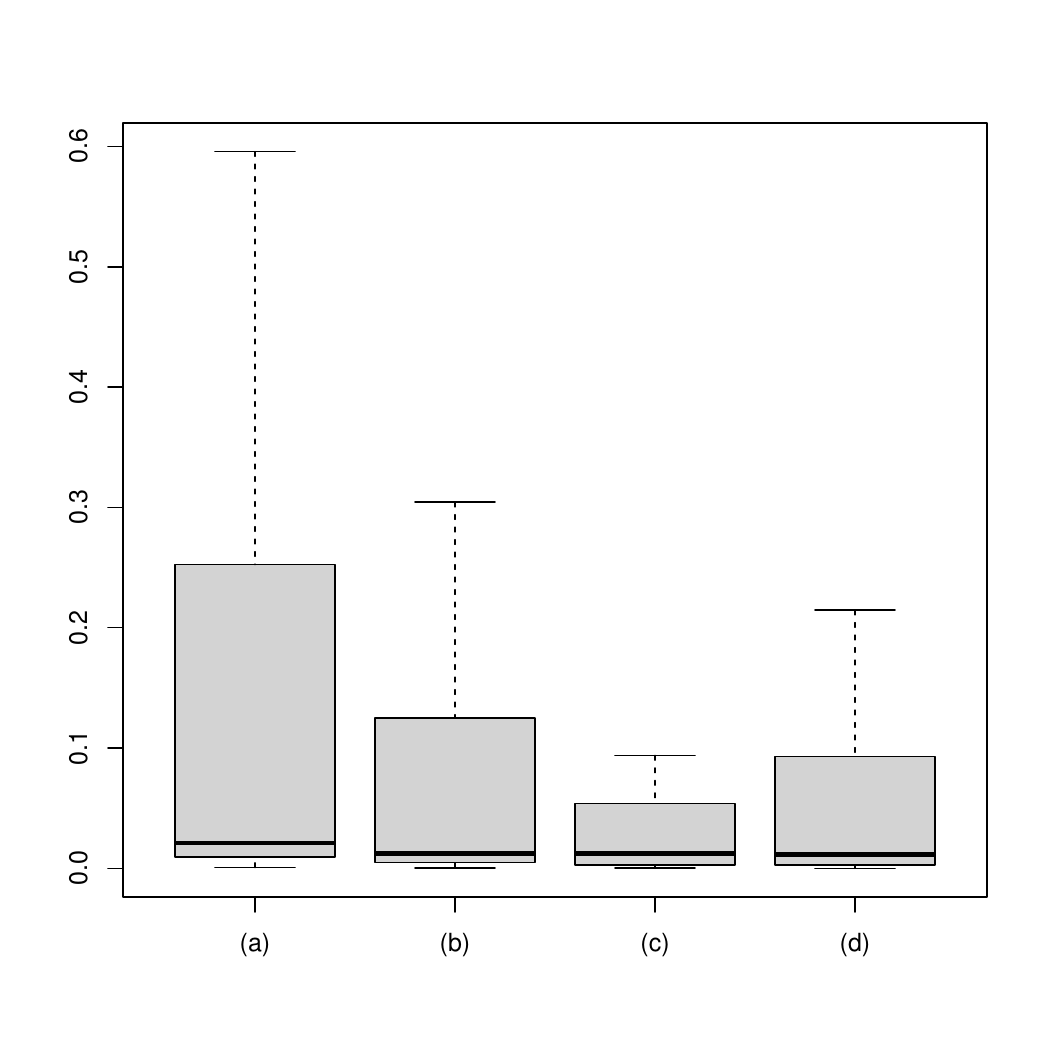}
\caption[FIGURE 5]{Riboflavin data. 4-fold cross-validation to evaluate the performance of Method 1 and Method 2 present in Table \ref{DescAp2}.}
\label{CVAp2}
\end{figure}

When we use Method 1 and the \texttt{glmmLasso} and the \texttt{EMLMLasso} algorithms, respectively (see Figure \ref{CVAp2}-(a) and Figure \ref{CVAp2}-(c)),  the $\text{MSE}_y$ was smaller with the proposed algorithm. The same occurs with Method 2 (see Figure \ref{CVAp2}-(b) and Figure \ref{CVAp2}-(d)). It can also be seen that for both algorithms, the use of Method 2 resulted in a smaller $\text{MSE}_y$.

\section{Discussion}\label{sec:conclusions}

In this work, we propose a novel algorithm for variable selection in linear mixed models based on the EM algorithm and the Lasso penalty, where the Lasso estimation step depends on \textsf{R} package \textbf{glmnet}. We call the proposed algorithm \texttt{EMLMLasso}. Even though other complex solutions have been proposed to deal with variable selection problems in linear mixed models, under low or high-dimensional settings, to the best of our knowledge, it is the first attempt to propose a straightforward implementation relying on existing packages. We focus on the Lasso penalty, but it certainly can be implemented for other kinds of penalties, such as ridge and elastic net.  We provide a publicly available \textsf{R}code to compute the methods introduced in this  paper, which is available for download from \texttt{GitHub}.

For comparison purposes,  we chose to use the publicly available \textsf{R} package \textbf{glmmLasso} \citep{Groll2022}, which is a well-known package for variable selection in generalized mixed-effects models. Under three scenarios, we investigate the performance of the proposed algorithm to select significant fixed effects through a set of simulations.  In the first scenario, we simulated covariates from the normal distribution and evaluated the capability of the \texttt{EMLMLasso} and \texttt{glmmLasso} algorithms to select the fixed effects. In a second scenario, we evaluated the ability of the algorithms to select fixed effects in the presence of categorical covariates. In a third scenario, we consider a large vector of fixed effects and evaluate the sensitivity and specificity of the algorithms. Finally, we use 10-fold cross-validation to evaluate the performance of algorithms under a high-dimensional setting ($p > n$). The results of the simulations demonstrated good properties of the proposed variable selection procedure. The \texttt{EMLMLasso} algorithm outperformed \texttt{glmmLasso} in all scenarios. Especially when evaluating the specificity, the proposed algorithm also stood out, even under a high-dimensional configuration.

We also analyzed two real data. The first is the Framingham heart study ($p<n$), with three covariates and all possible interactions, where the selected variables explain the cholesterol level, which is a risk factor for the evolution of cardiovascular diseases. In this first study, we added three more simulated variables: two are normal bivariates (evaluating the effect of correlated variables), and the other is Bernoulli (evaluating the effect of a categorical variable). The \texttt{EMLMLasso} selected some variables, and these were compatible with the results of the \textsf{R} packages \textbf{lme4} and \textbf{lmerTest}. However, \texttt{glmmLasso} selected the simulated covariates, and the results of the selected covariates were not compatible with the results provided by the same \textsf{R} packages. The second study is a gene expression data ($p>n$), where we are interested in relevant genes responsible for increasing the production of the riboflavin (vitamin B2) of \textit{bacillus subtilis}, a bacterium found in the human digestive tract. In this second study, as the covariates are correlated and $p>n$, we evaluated the two algorithms by adopting two configurations: 1) considering the original data and 2) using a function from \textsf{R} to eliminate linear correlations. The \texttt{EMLMLasso} made the selection of genes under the two considered strategies and presented a lower mean squared error for $\textbf{y}$.

The algorithm developed here does not consider censoring and/or missing responses, a typical problem in longitudinal studies. \cite{matos2013influence} have proposed a likelihood-based treatment based on the EM algorithm for parameter estimation in linear and nonlinear mixed-effects models with censored data (LMEC/NLMEC). Therefore, it would be a worthwhile task to investigate the applicability of variable selection in the context of LMEC/NLMEC models. Variable selection under skewness of the random effects \citep{Lachos:2010} is also a topic of our future research.


\section*{Acknowledgements} 
{The research of Daniela Carine Ramires de Oliveira was supported by Grant no. 401418/2022-7 from Conselho Nacional de Desenvolvimento Científico e Tecnologico (CNPq) – Brazil. Victor H. Lachos acknowledges the partial financial support from  UConn - CLAS's Summer Research Funding Initiative 2023.}



\bibliographystyle{chicago} 

\end{document}